\documentclass[preprint,showpacs,preprintnumbers,amsmath,amssymb,floatfix]{revtex4-1}
\usepackage{color}
\usepackage{graphicx}
\usepackage{dcolumn}
\usepackage{bm}
\usepackage{multirow}

\definecolor{sienna}{cmyk}{0,0.72,1,0.45}
\definecolor{fg}{cmyk}{0.91,0,0.88,.12}
\definecolor{yellow}{cmyk}{0,0,1,0}
\definecolor{or}{cmyk}{0,1,0.5,0}
\definecolor{magenta}{cmyk}{0,1,0,0}
\definecolor{rubinered}{cmyk}{0,1,0.13,0.45}
\definecolor{blue}{cmyk}{1,1,0,0}
\definecolor{turquoise}{cmyk}{1,1,0,0.5}
\definecolor{aquamarine}{cmyk}{0,1,0,0.0}
\definecolor{midnightblue}{cmyk}{1,0.5,0.0,0.0}
\definecolor{junglegreen}{cmyk}{1,0,0.2,0.5}

\begin{document}

\title{Time delay enhanced synchronization in a star network of second order Kuramoto oscillators}

\author{Ajay Deep Kachhvah}
\affiliation{Institute for Plasma Research, Bhat, Gandhinagar 382 428, India.}
\author{Abhijit Sen}
\affiliation{Institute for Plasma Research, Bhat, Gandhinagar 382 428, India.}

\begin{abstract}
We examine the onset of synchronization transition in a star network of Kuramoto phase oscillators in the presence of inertia and a time delay in the coupling. A direct correlation between the natural frequencies of the oscillators and their degrees is assumed. The presence of time delay is seen to enhance the onset of first order synchronization. The star network also exhibits different synchronization transitions depending on the value of time delay. An analytical prediction to observe the effect of the time delay is provided and further supported by simulation results. Our findings may help provide valuable insights into the understanding of mechanisms that lead to synchronization on complex networks.

\pacs{89.75.Hc}
\end{abstract}

\maketitle

\section{Introduction}

\noindent The dynamical features of a network of coupled oscillators provide a useful paradigm for understanding such diverse processes as epidemic spreading \cite{barrat}, traffic congestion \cite{barrat, bocca} and the general phenomenon of synchronization \cite{pecora, arenas}. Synchronization is ubiquitous in nature as well as in experimental systems studied in various branches of science and engineering \cite{piko,strog,nadis}. The presence of synchronization and its effects have been observed in power grids, communication networks, social interactions, circadian rhythms, and ecology \cite{arenas}. One of the important factors that influences synchronization behavior is the topology defining the pattern of connectivity of the underlying network \cite{moreno,lee,arenas2,zhou,garde}. The topology can also have an effect on the stability of the synchronized state \cite{pecora,bara,nishi,zhou2}. Among the large number of studies that have been carried out on synchronization in networks, the phenomenon of `explosive' synchronization has received important recent attention \cite{gardenes,leyva,peng}. The onset of such a synchronization is characterized by a discontinuous jump in the order parameter signifying a first order phase transition instead of the widely observed second order phase transition in the classical system of coupled phase oscillators introduced by Kuramoto \cite{kuramoto,kuramoto2}. 
The simple Kuramoto model of coupled phase oscillators neglects two 
important physical effects, namely, inertia in the intrinsic dynamics of the oscillators and finite time delay in the coupling mechanism.
Inertia introduces a second order time derivative in the dynamical equation of the individual oscillators and a coupled system of such oscillators is known to display distinctly different behavior from the usual first order Kuramoto system \cite{tanaka, peng, rohd, ace, dorf, trees}. For example, as shown by Tanaka \cite{tanaka}, the synchronization transition in a globally coupled network of Kuramoto oscillators with inertia is no longer second order but displays the characteristics of a first order phase transition. The presence of time delay in the coupling that arises naturally in any physical system due to the finite propagation speed of signals, also has a profound effect on the collective dynamics of the oscillator system \cite{choi,yeung,ares,herr,peron}.

In the present work our objective is to study the combined effect of inertia and time delay on the synchronization dynamics of a coupled system of Kuramoto oscillators in a scale-free network. To model the scale free network we choose a
star network as an approximation or as representative of the smallest unit of a scale-free network. We also assume a direct correlation between the natural frequencies of the oscillators and their degrees. In the absence of time delay the star network exhibits a discontinuous transition of synchronization as expected of a scale-free network. The inclusion of time delay brings about a rich variety in the collective behavior of this system.  First of all it introduces multiple synchronous states with attendant multi-stability in the network. Thus for a given coupling strength more than one stable state is now possible. A given synchronous state also exhibits frequency suppression as a function of the time delay. 
Each value of the time delay therefore presents a different synchronization transition in the star network, and by tuning the time delay we find that a desired phase transition can also be obtained. The time delay dependence of the average frequency is the key factor that leads to different synchronization transitions. We obtain our results from extensive numerical simulations carried out over a large range of parameters by using the package XPPAUT \cite{xpp}. We also provide analytical predictions for the time delay dependence of the average frequency of the system that are in good agreement with the simulation results.

\section{Model network and effect of time-delay}

\noindent Our model system consists of a star network that links $N$ second order Kuramoto phase oscillators through time delayed couplings. 
The model equations take the form,
\begin{equation}
m\ddot{\theta}_i + \dot{\theta}_i=\Omega_i + \lambda\sum_{j=1}^{N}
A_{ij}\sin[\theta_j(t-\tau)-\theta_i(t)],
\label{km2}
\end{equation}
where $\theta_i(t) (i=1,...,N)$ are the phases of the oscillators, $m$ is the mass, $\lambda$ is the homogeneous coupling strength, $\tau$ is the time delay in the coupling and $\Omega_i$ is the natural frequency of the $ith$ oscillator. The frequencies are selected from a given probability density $g(\omega)$. $A_{ij}$ are the elements of the adjacency matrix of connectivity of the network such that if two nodes $i$ and $j$ are connected then $A_{ij}=1$, otherwise $A_{ij}=0$. To track the synchronization transition effectively it is useful to look at the behavior of an order parameter defined as, 
\begin{equation}
re^{\iota\psi(t)}=\frac{1}{N}\sum_{j=1}^{N}e^{\iota\theta_j(t-\tau)},
\label{opar}
\end{equation}
where $\psi(t)$ represents an average phase of the collective dynamics of the system and the parameter $r$ provides a measure of the coherence of the collective motion of the oscillators or in other words the degree of synchronization among the oscillators. When the system is fully synchronized $r=1$, while $r=0$ denotes total incoherence. 
\begin{figure}[htb]
\begin{center}
\begin{tabular}{cccc}
(a)&
\hspace{-0.3cm}
\includegraphics[height=6cm,width=7.5cm]{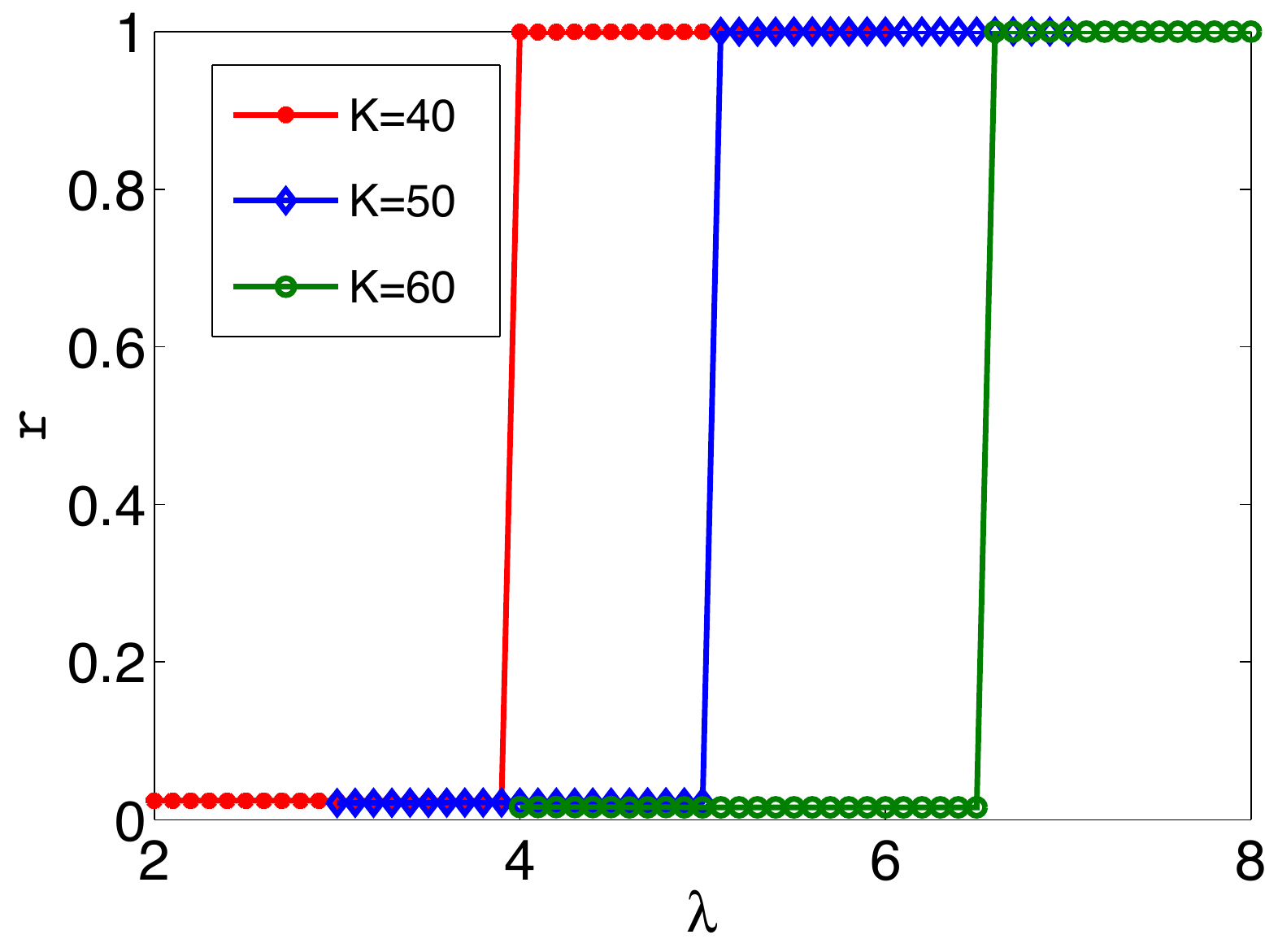}&
(b)&
\hspace{-0.3cm}
\includegraphics[height=6cm,width=7.5cm]{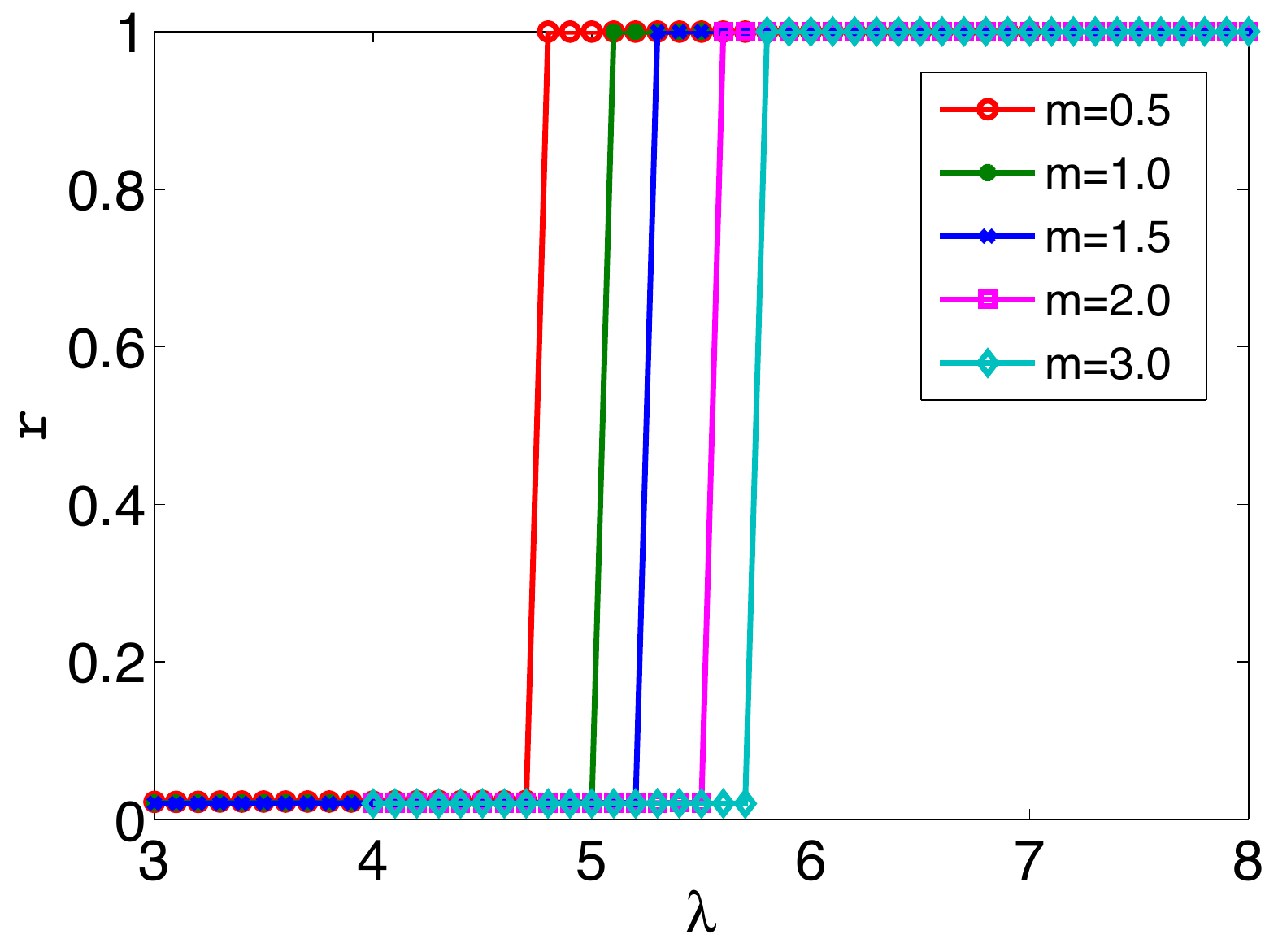}
\end{tabular}{}
\caption{(Color online) (a) Synchronization diagram $(r-\lambda)$ for the star networks of different sizes $K=40,50$ and $60$ corresponding to $m=1$. First order transition is observed in the star network when degree - frequency correlation is taken into account. (b) Synchronization diagram $(r-\lambda)$ for a  star network of $K=50$ leaves for different values of $m$.}
\label{fig:star_size}
\end{center}
\end{figure}

In order to locally investigate the effect of time-delay in a scale-free network, we consider an undirected star network. A scale-free network whose average degree is small can be represented as a collection of star networks because hubs would be more dominant. Therefore, here the hubs are represented as star motifs or networks.
In the star network a central hub is connected to $K$ leaves or peripheral nodes. Therefore, in a star network of $N=K+1$ nodes, the central hub has degree $k_h=K$ and each leaf or peripheral node has degree $k_l=1$. The natural frequency of each node is set equal to its respective degree, i.e., $\Omega_h=K$ for the hub and $\Omega_{l}=1$ for the peripheral nodes. We move to a rotating frame of the average phase $\psi(t)$ with frequency $\Omega$. In this frame the new variables for the hub and the leaves are then defined as $\phi_h=\theta_h-\Omega t$ and $\phi_i=\theta_i-\Omega t$, respectively. Using these definitions we can rewrite Eq.(\ref{km2}) for the hub and the leaves separately as, 
\begin{equation}
m\ddot{\phi_h}+\dot{\phi}_h=(\Omega_h-C) + \lambda\sum_{i=1}^{K}\sin[\phi_i(t-\tau)-\phi_h(t)-\Omega\tau],
\label{km2_hub}
\end{equation}
\begin{equation}
m\ddot{\phi_i}+\dot{\phi}_i=(\Omega_l-C) + \lambda\sin[\phi_h(t-\tau)-\phi_i(\tau)-\Omega\tau],
\label{km2_leaf}
\end{equation}
where the parameter $C=m\dot{\Omega}+\Omega$ is a function of the average frequency $\Omega$ of the system.
The global order parameter defined in Eq.(\ref{opar}) can be rewritten for the star network as following
\begin{equation}
re^{i\Psi(t)}=\frac{e^{i\theta_h(t)}+\sum_{i=1}^{K} e^{i\theta_i(t-\tau)}}{(K+1)}
\label{op_star}
\end{equation}
Multiplying both sides of Eq.(\ref{op_star}) by $e^{-i\theta_h(t)}$, taking the imaginary part, and writing in terms of phases $\phi_h$ and $\phi_i$, Eq.(\ref{km2_hub}) can then be rewritten as
\begin{equation}
m\ddot{\phi_h}+\dot{\phi_h}=(\Omega_h-C) - \lambda(K+1)r\sin\phi_h(t).
\end{equation}
In the phase-locked condition all time derivatives vanish. Hence for the hub $\dot{\phi_h}=\ddot{\phi_h}=0$, which gives us,
\begin{equation}
\sin\phi_h = \frac{(\Omega_h-C)}{\lambda r(K+1)}.
\label{res_hub}
\end{equation}
Imposing a similar condition for the phase-locked peripheral nodes or leaves, i.e., $\dot{\phi_i}=\ddot{\phi_i}=0$, Eq.(\ref{km2_leaf}) reduces to
\begin{align}
\cos[\phi_i(t)+\Omega\tau]=\frac{(C-\Omega_l)}{\lambda}\sin\phi_h(t-\tau) 
\pm\frac{1}{\lambda}\sqrt{[\lambda^2-(\Omega_l-C)^2][1-\sin^2\phi_h(t-\tau)]}.
\label{res_leaf}
\end{align}
At the critical coupling $\lambda=\lambda_c$, from Eq. (\ref{res_leaf}) we obtain $\cos[\phi_i(t)+\Omega\tau]=\sin\phi_h(t-\tau)$, which further leads to $\phi_h(t-\tau)-\phi_i(t)=\frac{\pi}{2}+\tau\Omega(\tau)$. Therefore, we can set the time delay in order to obtain different phase couplings between the hub and the peripheral nodes.
Eq.(\ref{res_leaf}) is valid only for $\lambda\geq|\Omega_l-C|$, so that the values of the critical coupling for the existence of synchronized state would be determined by $\lambda_c=|\Omega_l-C|$. Therefore, $\lambda\geq\lambda_c$ is a necessary condition for the existence of the synchronous solutions. Since the value of the average frequency depends on the time delay, i.e. $\Omega=\Omega(\tau)$, the value of the critical coupling $\lambda_c$ also depends on the time delay $\tau$. 

For $\tau=0$ our above model system 
exhibits discontinuous synchronization transition as shown in Fig.\ref{fig:star_size}(a). The order parameter $r$ is seen to make a discontinuous jump at a critical value of the coupling strength $\lambda$. This behavior is seen for different sizes of the network ranging over $K=40,50$ and $60$, and it is observed that the critical value of the coupling strength increases with the size of the network. This is understandable since for a larger size of the network a larger number of oscillators have to be entrained to 
obtain synchronous behavior. We should also mention that the model star graph studied here should be viewed as an approximation to a larger scale-free network with a distribution of hubs some of which would have a large number of links. The single hub model is representative of one such hub and the values of $K$ chosen are representative of the expected degrees for such a hub in a realistic  scale-free network with $N=500$ to $1000$ nodes. In Fig.\ref{fig:star_size}(b) we show the effect of inertia (still with $\tau=0$) on
the critical coupling for synchronization for a given network size ($K=50$) by varying the value of the mass $m$. As can be seen an increase in $m$ increases the threshold value for the onset of synchronization. However the nature of synchronization, namely a first order phase transition, does not change.

\begin{figure}[htb]
\begin{center}
\includegraphics[height=6cm,width=7.5cm]{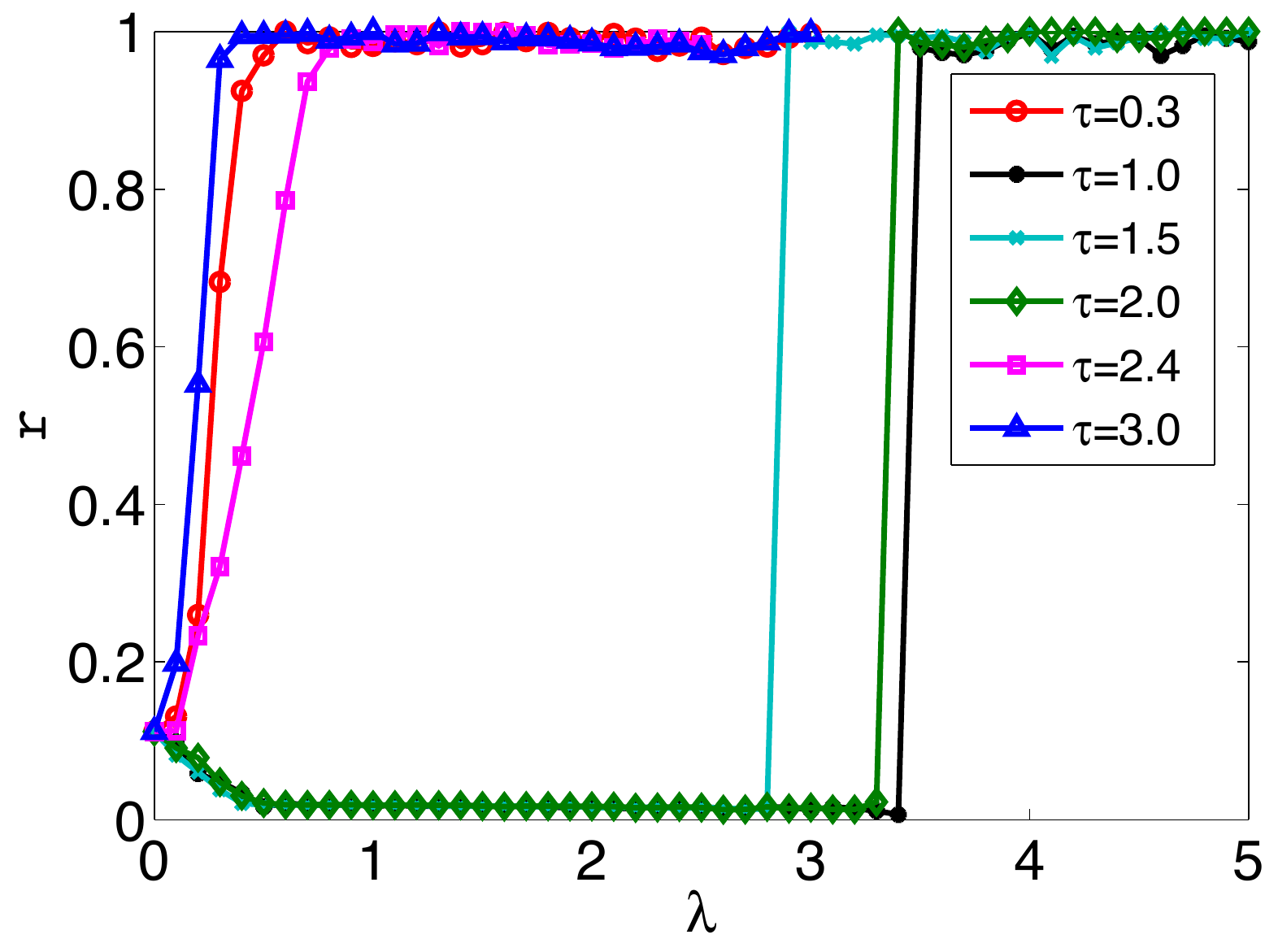}\\
\caption{(Color online) The order parameter $r$ as a function of coupling strength $\lambda$ with different values of time-delay $\tau$ introduced in a star network of $K=50$ leaves with $m=1$. For different values of time delay, the star network represents different synchronization transition.}
\label{fig:star_td}
\end{center}
\end{figure}

We next look at the behavior of the system in the presence of time delay. We find that for finite values of time delay, the synchronization behavior of the system undergoes significant changes and this is shown in 
Fig.\ref{fig:star_td} where the order parameter is plotted against $\lambda$ for a fixed value of $K=50$ and $m=1$. We observe that the presence of a time delay can lower the threshold for phase transition to a synchronous state compared to that in the absence of time delay. For instance, the value of critical coupling $\lambda_c=5$ for $\tau=0$ is now lowered to $\lambda_c=2.8$ when $\tau=1.5$. Fig.\ref{fig:star_td} also demonstrates that each value of the time delay gives rise to a different synchronization transition, namely a first order or second order phase transition. For  example, for $\tau=1.5$ a first order, and for $\tau=2.4$ a second order phase transition is obtained.

\begin{figure}[htb]
\begin{center}
\includegraphics[height=6.5cm,width=7.5cm]{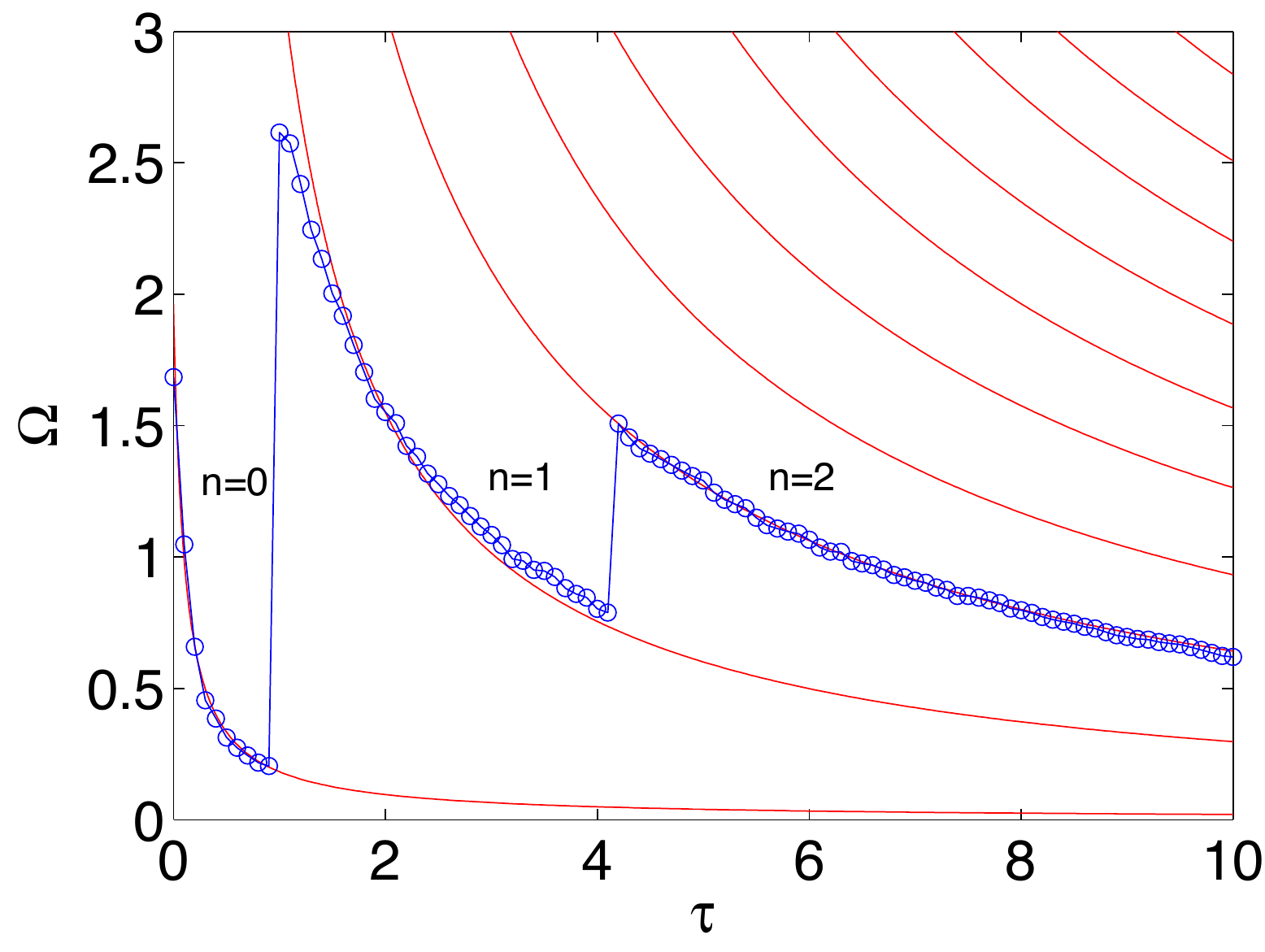}\\
\caption{(Color online) The average frequency $\Omega$ (or $C$) as a function of the time delay $\tau$ for the coupling strength $\lambda=5$ in a star network of $K=50$ leaves. The circles $\circ$ represent the simulation result, and solid lines represent the theoretical prediction by Eq.(\ref{fq_del}) while considering $\Omega\tau\rightarrow(\Omega\tau-n\pi)$. There is an oscillatory dependence of $\Omega$ on the time delay $\tau$.}
\label{fig:omg_tau}
\end{center}
\end{figure}

Such interesting behavior results from a dependence of the average frequency $\Omega$ or the parameter $C$ on the time delay $\tau$, i.e., $\Omega=\Omega(\tau)$. Since the critical coupling $\lambda_c$ depends on the parameter $C\approx\Omega$, therefore it is fruitful to analyze how the parameter $C\approx\Omega$ varies with the time delay $\tau$. In Fig.\ref{fig:omg_tau} the average frequency $\Omega$ is shown as a function of time delay $\tau$. One notes two important characteristic features about the dependence of $\Omega$ on $\tau$. Starting from $\tau=0$ as one increases $\tau$ the average frequency $\Omega$ starts decreasing displaying the so-called frequency suppression phenomenon that has been noted before for time delayed systems \cite{choi}. Beyond a certain value of $\tau$ the frequency $\Omega$ jumps to a higher
value and continues to decrease again with $\tau$ along another branch. This behaviour is repeated as one moves along to higher values of $\tau$ - indicating an oscillatory pattern. This oscillatory dependence between $\Omega$ and $\tau$ accounts for the different behavior of the synchronization transition exhibited by the star network (see Fig. \ref{fig:star_td}) in contrast to the case without time delay where one only observes a first order transition (see Fig. \ref{fig:star_size}).

\begin{figure}[htb]
\begin{center}
\includegraphics[height=6cm,width=7.5cm]{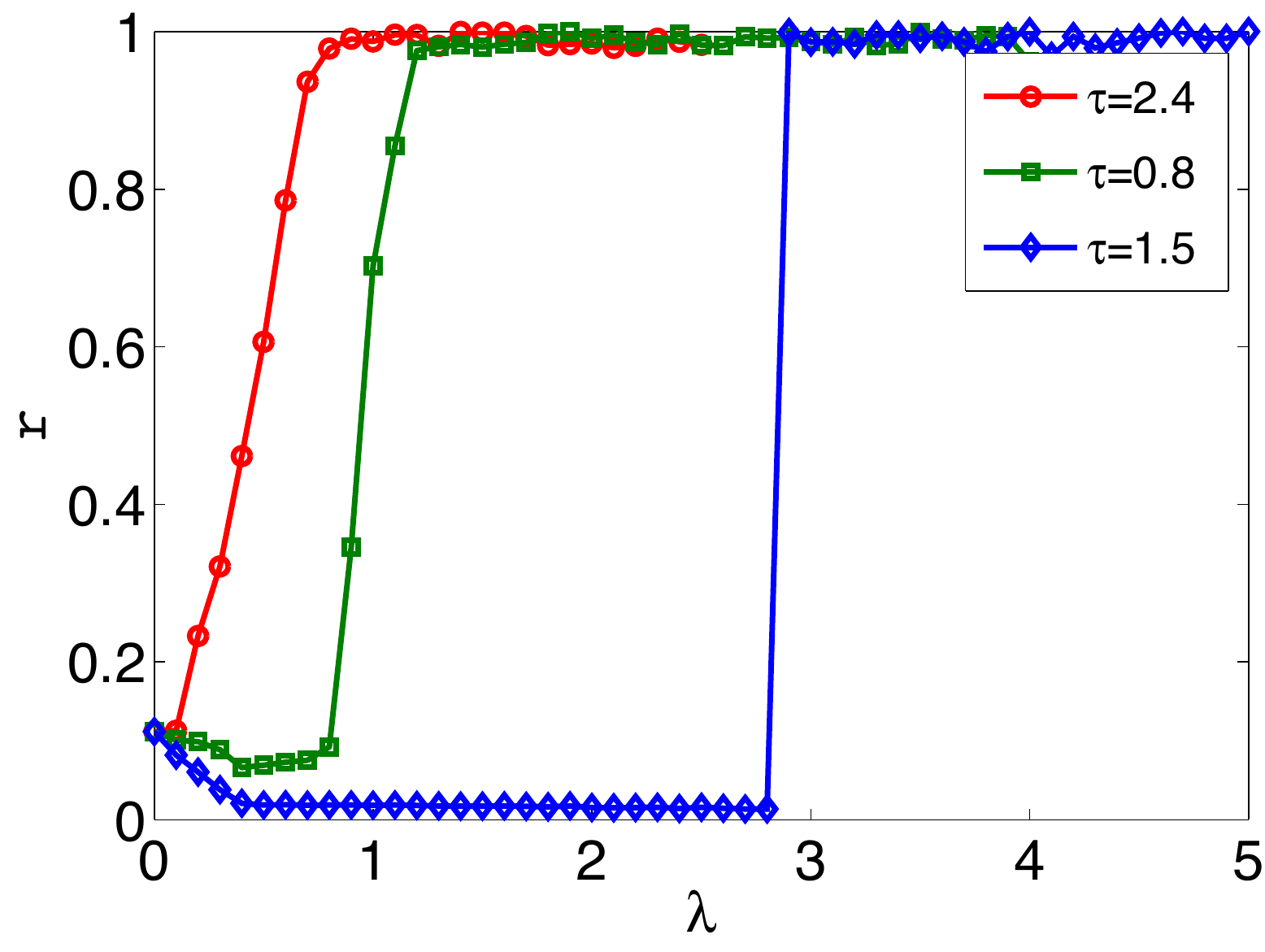}\\
\caption{(Color online) The order parameter $r$ as a function of coupling strength $\lambda$ for the time-delay $\tau=2.4,0.8$ and $1.5$ present in a star network of $K=50$ nodes.}
\label{fig:omgtau_cmp}
\end{center}
\end{figure}

For a given oscillatory dependence between $C\approx\Omega$ and $\tau$, we can choose the time delay $\tau$ which makes the system prone to the onset of synchronization. For a star network of given size $K$, we can find out a value for the time delay that enhances the synchronization transition level. A value of the time delay can be determined which would produce the maximum value of the critical coupling. Since we know the critical coupling $\lambda_c=|\Omega_l-C|$, therefore the lower differences between $\Omega_l$ and $C$ yields lower value of $\lambda_c$, which enhances the onset of synchronization. By contrast, higher frequency differences would yield higher value of $\lambda_c$ and disturb the onset of synchronization. These cases are supported by  synchronization diagram in Fig.\ref{fig:omgtau_cmp} for the values of the time delay chosen according to the oscillatory dependence between $\Omega$ and $\tau$ from Fig. \ref{fig:omg_tau}. For instance, for the time delay of $\tau=1.5$, the difference $|\Omega_l-C|$ is higher giving rise to a first order transition. However, for a time delay of $\tau=2.4$, the difference $|\Omega_l-C|$ is low which leads to a second order transition.
It is found that a first order transition is observed if the time delay is set to obtain a higher value of the critical coupling $\lambda_c$. 

In the synchronized state, there is an abundance of synchronized nodes which are locked to the mean field that rotates with a constant frequency $\Omega$, so that $C\approx\Omega$.
In order to analyze the parameter $C$, we have assumed that after transition to synchrony all nodes in the star network are evolving with the same average phase $\psi$, i.e., $\theta_h=\psi$ for the hub and $\theta_i=\psi$ ($i=1,2,...,K$) for the peripheral nodes. Substituting these solutions in Eqs. (\ref{km2_hub}) and (\ref{km2_leaf}) and writing a set of $K+1$ equations, we obtain
\[
\begin{array}{cc}
 C=K-K\lambda\sin (\Omega\tau) & \mbox{for hub} \\
 C=1-\lambda\sin (\Omega\tau) & \mbox{for each leaf} 
\end{array}
\]
After summing up the above set of $(K+1)$ equations for the hub and $K$ peripheral nodes, we obtain the parameter $C\approx\Omega$ of the star network given by
\begin{equation}
C\approx\Omega=\frac{2K}{K+1}[1-\lambda\sin (\Omega\tau)].
\label{fq_del}
\end{equation}
Eq.(\ref{fq_del}) is similar to the relation between the average frequency $\Omega$ and the time delay $\tau$ of the star network of first order Kuramoto oscillators in presence of the time delay \cite{peron}. Therefore, in our system as well we expect a similar kind of oscillatory dependence between $\Omega$ and $\tau$ as seen in the case of Ref. \cite{peron}. From Ref. \cite{earl}, the necessary condition for the stability of the synchronous states ($\theta_h=\theta_i=\psi$) is given by $\cos(\Omega\tau)>0$.
Thus, the product $\Omega\tau$ must satisfy $2n\pi<\Omega\tau<(2n+1)\pi$ for $n=0,1,2,...$. Therefore, product $\Omega\tau$ should be replaced by ($\Omega\tau-n\pi$) in Eq. (\ref{fq_del}). Thus Eq. (\ref{fq_del}) when $\Omega\tau\rightarrow(\Omega\tau-n\pi$) is the analytical prediction for the average frequency $\Omega$ and the time delay $\tau$ for the star network. In Fig. \ref{fig:omg_tau}, Eq. (\ref{fq_del}) is plotted for $n=0,1,2...$, and $\lambda=5$ for a star network of $K=50$. From Fig. \ref{fig:omg_tau} it is obvious that there is a good agreement between the simulation results and the analytical predictions made by Eq. (\ref{fq_del}). The critical coupling $\lambda_c$ for the star network can also be determined by rewriting the Eq. (\ref{fq_del}) in terms of $\lambda_c$ and $\tau$ using the expression $\lambda_c=|\Omega_l-C|$ and considering $\Omega\tau\rightarrow(\Omega\tau-n\pi)$.   

\section{Conclusions}

\noindent To summarize, in this paper we have studied the synchronization behaviour of a scale-free network of coupled oscillators in the presence of inertia in the system and time delay in the coupling. The presence of inertia is modelled by using second order Kuramoto oscillators and the scale-free network is approximated by a star network and by setting the frequency of each node equal to its degree of connection. In the absence of time delay the network shows the onset of first order transitions to synchronized behaviour and the threshold for this transition increases with an increase in inertia. The presence of time delay can mitigate to some extent the influence of inertia by lowering the threshold for synchronization. In addition time delay can also influence the nature of the synchronization transition by making it switch from a first order to a second order transition. The mechanism underlying this change is associated with the change in the average frequency of the system as a function of time delay. Thus time delay and inertia provide us with parametric handles  that can be used to control the onset of synchronization in a scale-free network. Time delay offers a further facility of changing the nature of the transition from a first order to a second order synchronization. Our model calculations based on the star network agree very well with the numerical simulation results. Since both inertia and time delay are likely to be present in any realistic physical network our results can help in understanding the microscopic nature of synchronization phenomena in them and also provide a means of controlling the nature and onset of such synchronizations.

\begin{acknowledgments}
ADK thanks Gautam C. Sethia for useful discussions and also help towards using the software XPPAUT.

\end{acknowledgments}


\end{document}